\documentclass[PRD,reprint,onecolumn,showpacs,
 nofootinbib, superscriptaddress]{revtex4}

\usepackage{color}
\usepackage{amsmath,amssymb}
\usepackage[colorlinks,allcolors=blue]{hyperref}
\usepackage{graphicx}
\usepackage{dcolumn}
\usepackage{bm}
\usepackage{url}

\newcommand{\beq}{\begin{equation}}
\newcommand{\eeq}{\end{equation}}
\newcommand{\beqa}{\begin{eqnarray}}
\newcommand{\eeqa}{\end{eqnarray}}

\begin{document}

\title{On the morphology of $\gamma-$ray emission induced by $e^{\pm}$ from annihilating self-interacting dark matter}

\author{Ming-Yang Cui}
\affiliation{Department of Physics, Nanjing University, Nanjing 210093, P. R. China.}
\affiliation{Key Laboratory of Dark Matter and Space Astronomy, Purple Mountain Observatory, Chinese Academy of Sciences, Nanjing 210008, China}
\author{Cun Zhang}
\affiliation{Department of Physics, Nanjing University, Nanjing 210093, P. R. China.}
\affiliation{Key Laboratory of Dark Matter and Space Astronomy, Purple Mountain Observatory, Chinese Academy of Sciences, Nanjing 210008, China}
\author{Hong-Shi Zong}\email{zonghs@nju.edu.cn}
\affiliation{Department of Physics, Nanjing University, Nanjing 210093, P. R. China.}
\date{\today}

\begin{abstract}
With the Fermi-LAT data quite a few research groups have reported a spatially extended GeV $\gamma$-ray excess surrounding the Galactic Center (GC). The physical origin of such a GeV excess is still unclear and one interesting possibility is the inverse Compton scattering of the electrons/positrons from annihilation of self-interacting dark matter (SIDM) particles with the interstellar optical photons.
In this work we calculate the morphology of such a kind of $\gamma$-ray emission. For the annihilation channel of $\bar{\chi}\chi\rightarrow \phi\phi\rightarrow e^{+}e^{-}e^{+}e^{-}$, the inverse Compton scattering (ICS) dominates over the bremsstrahlung on producing the GeV $\gamma$-ray emission. For the SIDM particles with a rest mass $m_\chi \sim $ tens GeV that may be favored by the modeling of the Galactic GeV excess, the ICS radiation at GeV energies concentrates along the Galactic plane.
The degrees of asymmetry high up to $\geq 0.3$ are found in some regions of interest, which in turn proposes a plausible test on the SIDM interpretation of the GeV excess.
\end{abstract}


\maketitle

\section{Introduction}\label{sec:intro}
In the standard $\Lambda$CDM cosmology model, cold dark matter (DM) consists of $\sim 26.8\%$ energy density in the current universe \cite{Ade:2013aa}. Various well-motivated particle candidates have been proposed in the literature and the weakly interacting massive particles (WIMPs) are the most extensively discussed particles \cite{Jungman:1995df, Bertone:2004pz, Hooper:2007qk, Feng:2010gw, Fan:2010yq}. In principle,
WIMPs may be able to annihilate with each other (or alternatively decay) and produce energetic particle/antiparticle pairs and gamma-rays. Search for the dark matter originated signals in high energy cosmic rays and gamma-rays is one of the prime targets of some space missions such as PAMELA \footnote{\url{http://pamela.roma2.infn.it/index.php}}, Fermi-LAT \footnote{\url{http://fermi.gsfc.nasa.gov/}}, AMS-02 \footnote{\url{http://www.ams02.org/}} and DAMPE \footnote{\url{http://dpnc.unige.ch/dampe/}}.
Though abundant, so far the DM particles have not been reliably detected. Nevertheless, some tentative signals have attracted wide attention. Among them the most-widely examined signals include the electron/positron cosmic ray anomaly \cite{Adriani:2008zr, Aguilar:2013qda, Adriani:2011xv, Chang:2008aa, FermiLAT:2011ab,Li:2014csu} and the so-called Galactic center GeV excess, an ``unexpected" spatially extended GeV $\gamma$-ray emission surrounding the Galactic Center (GC) \cite{Goodenough:2009gk, Vitale:2009hr, Hooper:2010mq, Hooper:2010im, Abazajian:2012pn, Gordon:2013vta, Huang:2013pda,Hooper:2013rwa,Daylan:2014rsa,Zhou:2014lva, Calore:2014xka,Huang2015,FermiLAT:2015,Liang:2016}. Hereafter we denote the GC excess component as the GCE. The GCE is at GeV scale and extends to a Galactic latitude $|b| \sim 10^\circ-20^\circ$ \cite{Hooper:2013rwa}.
Both the spectrum and the morphology of the GCE are found to be compatible with that predicted from the annihilations of WIMPs with a rest mass $\sim$ tens GeV via the channels mainly to quarks \cite{Daylan:2014rsa}. The GCE has also been found to be robust across a variety of models for the diffuse galactic $\gamma-$ray emission \cite{Zhou:2014lva, Calore:2014xka,Huang2015,FermiLAT:2015}.

In the literature some models have been proposed for this peculiar signal in the GC, such as the millisecond pulsars (MSPs) \cite{Abazajian:2012pn,Mirabal2013, Yuan:2014rca, LeeMSP,Lee2015,Brandt:2015,Bartels:2015}, the leptonic cosmic ray outbursts \cite{Cholis_lepbur} and dark matter annihilation \cite{Hooper:2010mq, Hooper:2010im, Abazajian:2012pn, Gordon:2013vta, Huang:2013pda,Hooper:2013rwa,Daylan:2014rsa,Zhou:2014lva,Calore:2014xka}. One type of the DM model is the so-called  self-interacting dark matter (SIDM) \cite{Spergel2000, Yoshida:2000, Carlson1992}. Kaplinghat et al. \cite{Yu GeV} showed that a special SIDM model could account for the GCE. In such a model the SIDM particles interact with each other via a light mediator and annihilate to energetic $e^{+}e^{-}$ pairs.
Then the high energy $e^{\pm}$ scatter with the starlight and afterward boost the optical photons to higher energies that can be estimated as $\epsilon_{\rm IC} \sim 4\gamma_{e^\pm}^2 \epsilon_{\rm starlight}/3$, where $\gamma_{\rm e^\pm}$ is the Lorentz factor of the $e^\pm$ pairs formed in the dark matter annihilation and $\epsilon_{\rm starlight}\sim 1$ eV is the typical energy of the starlight. For the dark matter particles with a rest mass $m_\chi \sim 40$ GeV, in the case of $\bar{\chi}\chi\rightarrow \phi\phi\rightarrow e^{+}e^{-}e^{+}e^{-}$ we have $\gamma_{e^\pm} = m_{\chi}/2m_{\rm e} \approx 4\times 10^{4}~(m_\chi/40~{\rm GeV})$ and $\epsilon_{\rm IC} \sim 2~{\rm GeV}~ (m_\chi/40~{\rm GeV})^{2}(\epsilon_{\rm starlight}/1~{\rm eV})$, where $m_{\rm e}$ is the rest mass of the electrons/positrons.
A detailed numerical investigation demonstrates that the inverse Compton scattering (ICS) process can reasonably explain the GCE spectrum for $m_\chi \sim 20-60$ GeV \cite{Yu GeV}. Moreover, on one hand, numerical simulations have shown that nuclear-scale
dark matter self-interaction cross sections can produce heat transfer from the hot outer region to the cold inner region of dark matter halos, reducing the central densities of dwarf galaxies in accordance with observations \cite{Vogelsberger2012,Rocha2013,Zavala2013,Elbert:2014}. On the other hand, the $e^\pm$ pairs produced via dark matter annihilation will not produce plentiful $\gamma-$rays from
dwarf galaxies due to the hosted low starlight and gas densities, in agreement with the non-detection of the statistical significant gamma-ray signal from the spherical dwarf galaxies \cite{Ackermann:2015zua,Sameth2015, Li shang:2015}.

In this work, we investigate in detail the spatial distribution of the ICS component. The main concern is the following: For the prompt gamma-rays resulting in the final state radiation of dark matter annihilation, the morphology should be directly governed by the dark matter distribution and is hence expected to be spherically symmetric with respect to the GC. For the ICS component, however, the situation is more complicated since the distribution of the interstellar radiation field is not isotropic. The spatial distribution of the ICS component does not follow the electron/positron pairs originated from dark matter annihilation unless the electrons/positrons are energetic enough to lose most of their energy quickly via ICS. {  The morphological properties of inverse Compton emission from DM annihilations have been investigated previously in the literature \cite{Regis2009,Borriello2009,Dobler2010,Dobler2011,Yuan:2015}. Herein, we propose a new method to quantitatively calculate the asymmetry of spatial distribution of gamma-ray skymap and subsequently apply it to the specific SIDM model for the GCE.} In this work we adopt the GALPROP v54 code  \footnote{\url{http://galprop.stanford.edu}} \cite{Strong:1998pw} to numerically calculate the morphology of the ICS component.

This work is structured as follows. In Sec. \ref{Sec:calculation} we briefly introduce the calculation of the spatial distribution of diffuse $\gamma$-ray emission of DM annihilations and the regions of interest (ROI) adopted in this paper. In Sec. \ref{Sec:analysis} we calculate the degree of ``departure" from the rotational-symmetry of the morphology of $\gamma$-ray emission in two ways.
We take bremsstrahlung and prompt emission into account and test the influence of Cosmic Ray Propagation Parameters on numerical analysis in Sec. \ref{sec:IIIC}.
In Sec. \ref{Sec:discussion} we summarize our results with some discussion on the prospect of testing the SIDM annihilations origin of the GCE.


\section{The spatial distribution of diffuse $\gamma$ rays emission and our regions of interest}\label{Sec:calculation}
\begin{figure}
    \begin{center}
        \includegraphics[width=0.44\linewidth]{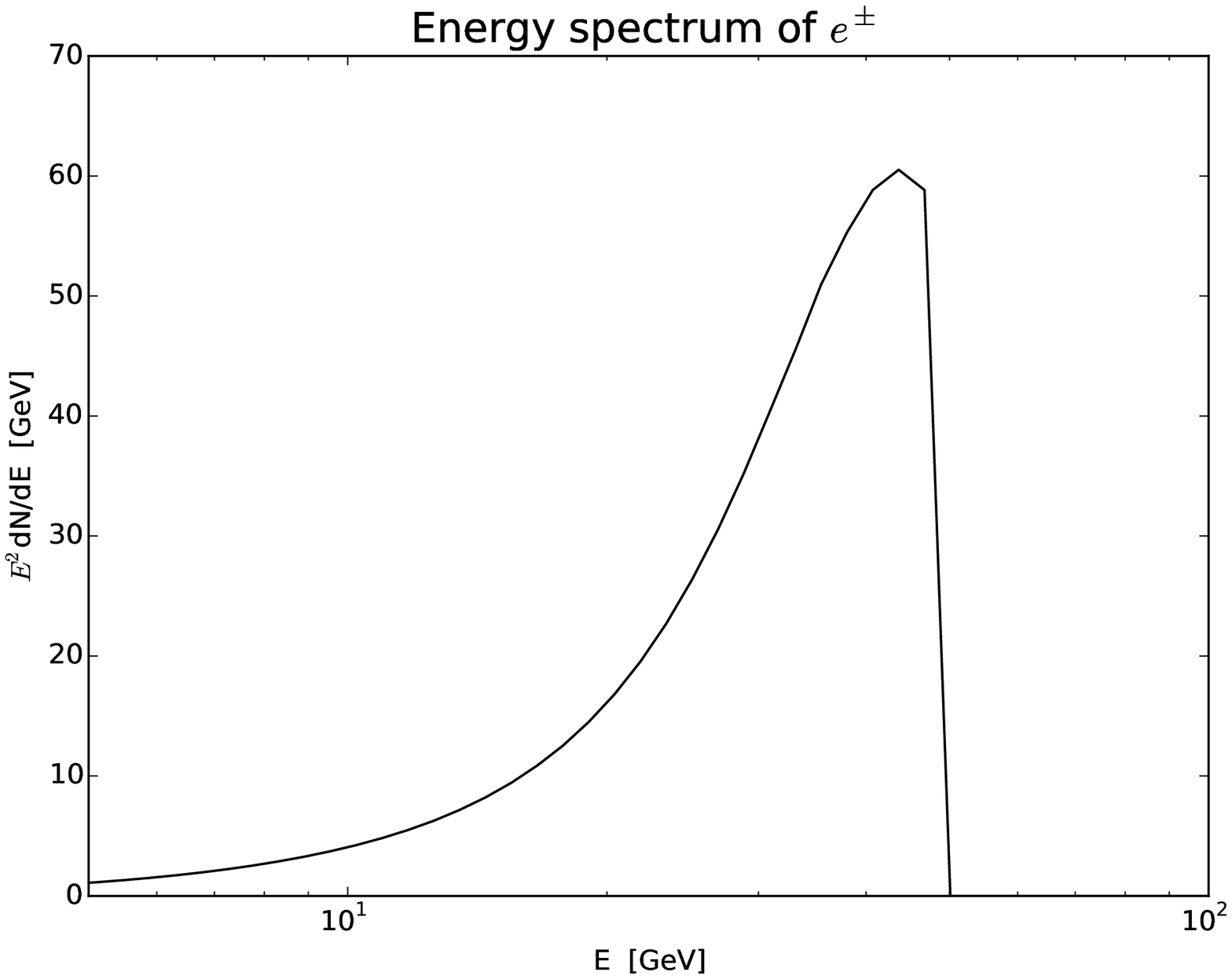}
        \includegraphics[width=0.46\linewidth]{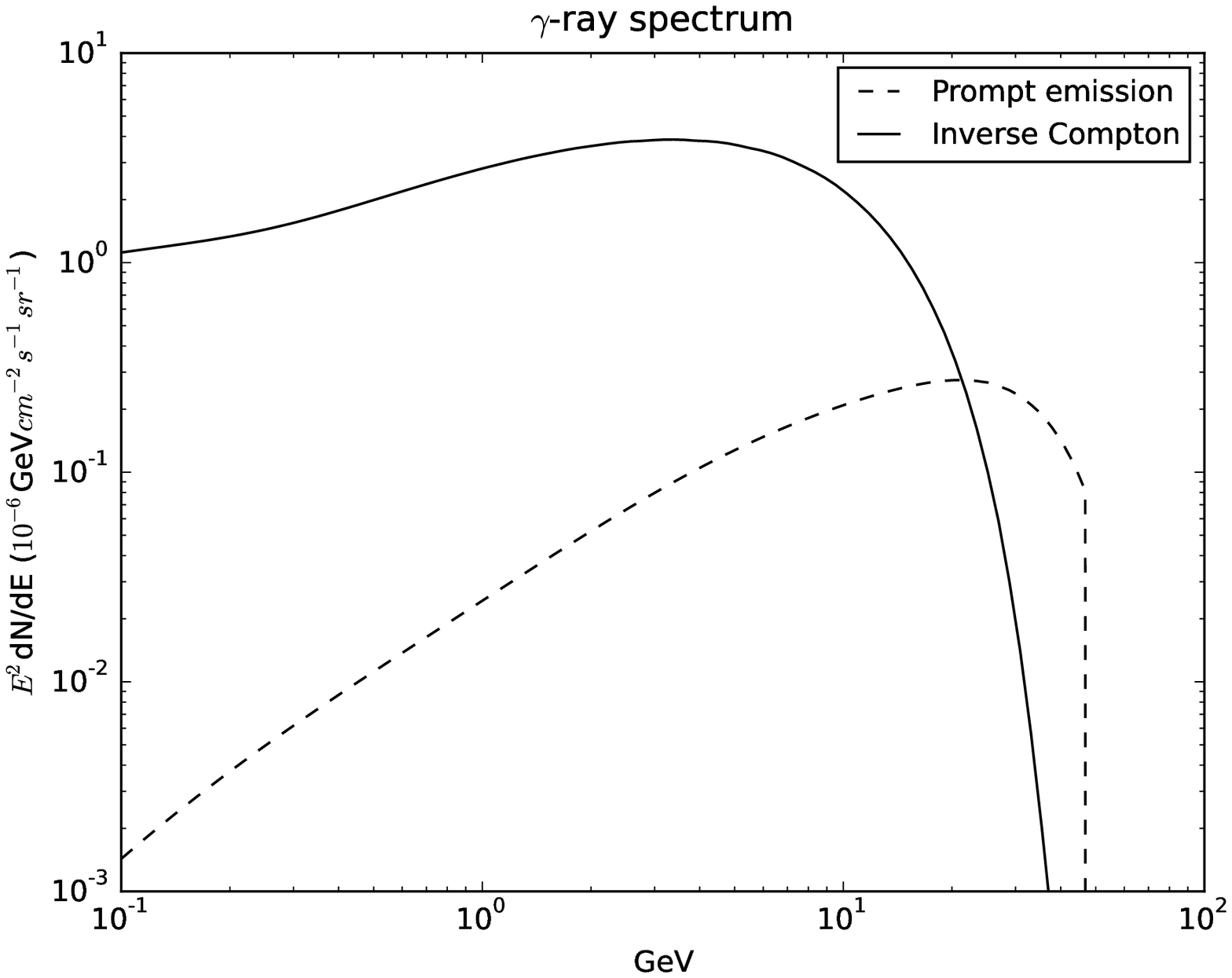}
    \end{center}
\caption{{  The left panel: the energy spectrum of $e^{\pm}$ resulting in $\chi\chi\rightarrow \phi\phi\rightarrow e^{+}e^{-}e^{+}e^{-}$ for $m_\chi =50$ GeV. The right panel: the prompt (i.e., the final state radiation of DM annihilations) and the ICS spectra, both are averaged in the regions of $|l| <5^\circ$ and $2^\circ< |b| <5^\circ$.}}
\label{fig:spectrum}
\end{figure}

The cosmic ray electrons/positrons propagating through the Milky Way interact with interstellar gas, magnetic fields as well as the interstellar radiation field (ISRF \cite{Porter:2005}) and generate high energy gamma-ray emission.
In this work we consider the electron/positron pairs only resulting from SIDM annihilation. The spectra of such electron/positron pairs are calculated with the software PPPC4DMID \cite{PPPC} {  (see the left panel of Fig. \ref{fig:spectrum})}. The dark matter distribution is taken to be the generalized NFW profile $\rho \propto (r/r_{\rm s})^{-\alpha}(1+r/r_{\rm s})^{-3+\alpha}$ \cite{Navarro:1995iw, Navarro:1996gj}, where $r_{\rm s}=20$ kpc is the scale radius and  $\alpha=1.2$ is the slope index (such an $\alpha$ was favored in the modeling of the GCE \cite{Hooper:2013rwa,Daylan:2014rsa}). The local DM energy density is taken to be  $0.3~{\rm GeV ~ cm^{-3}}$. For illustration we take $m_\chi =50$ GeV and $\langle\sigma v\rangle=6\times10^{-26} ~{\rm cm^{3}~s^{-1}}$ except in Sec. \ref{sec:IIIC} where we will also discuss the case of $m_\chi =1$ TeV.

We take the GALPROP v54 code \cite{Strong:1998pw} to calculate the propagation of $e^{\pm}$ and $\gamma$-ray emission through ICS as well as bremsstrahlung. The diffusion with re-acceleration configuration of the propagation model is adopted.
And we take one group of propagation parameters as default parameters in the bulk of this paper for simplicity except in Sec. \ref{sec:IIIC} where two different propagation parameters would be used for comparison. The default parameters comprise diffusion coefficient $D_{0}$=5.3$\times10^{28}$ $cm^{2}$ $s^{-1}$ at the reference rigidity $R = 4$ GV, the height of the propagation halo $z_{h}=4$ kpc, the Alfven speed $v_{A}=33.5~{\rm km ~s^{-1}}$ which characterizes the re-acceleration, and the power-law index $\delta$=0.33 of the rigidity dependence of the diffusion coefficient \cite{Trotta_CR,Jin_CR}.

\begin{figure}
    \begin{center}
        \includegraphics[width=0.45\linewidth]{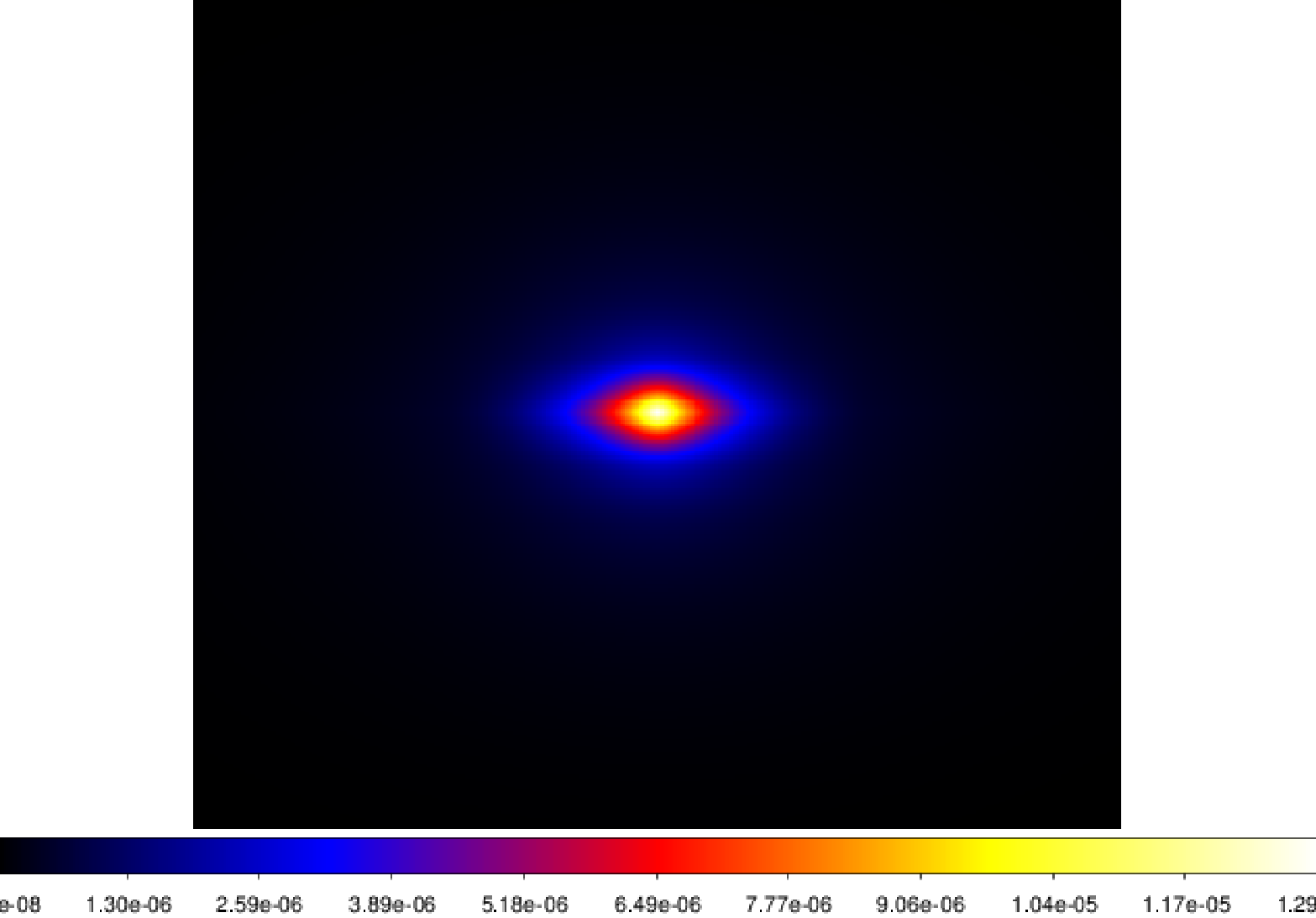}
        \includegraphics[width=0.45\linewidth]{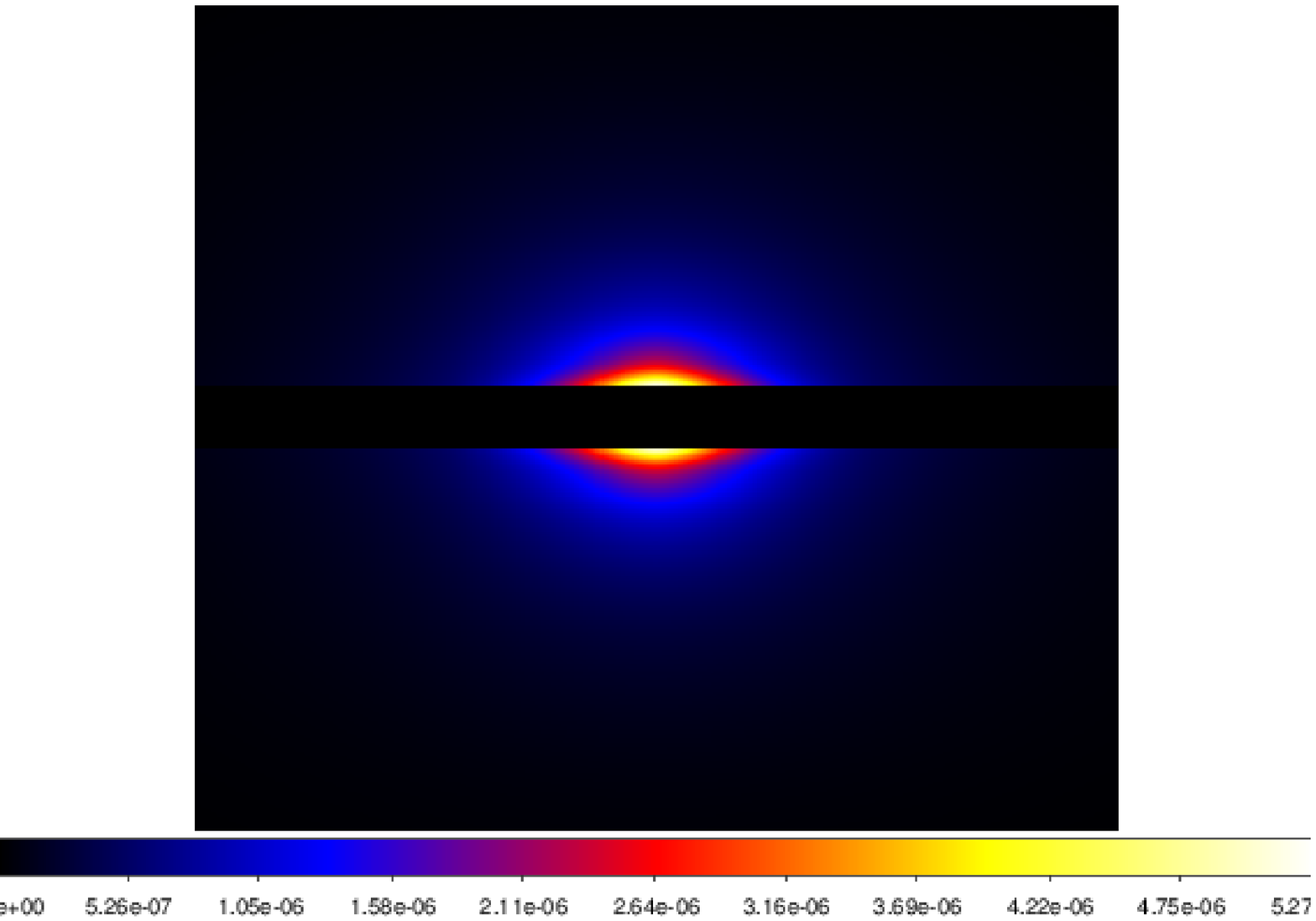}
    \end{center}
\caption{The count maps of ICS $\gamma$-ray (in the energy range of $1~{\rm GeV}-3.16~{\rm GeV}$) are shown in the regions of $|b| <30^\circ$ and $|l| <30^\circ$ and in unit of ${\rm photons~cm^{-2}~s^{-1}~sr^{-1}}$. In the left panel the regions of $|b|<2^\circ$ are kept while in the right panel such a region is masked.}
\label{fig:countmap}
\end{figure}

We focus on the ICS $\gamma$ rays in the energy range of $1~{\rm GeV}-3.16~{\rm GeV}$ in which the GCE likely peaks \cite{Daylan:2014rsa,Zhou:2014lva, Calore:2014xka,Huang2015,FermiLAT:2015}. The result is shown in the left panel of Fig. \ref{fig:countmap}, covering regions of $|b|<30^\circ$ and $|l| <30^\circ$, in unit of ${\rm photons~cm^{-2}~s^{-1}~sr^{-1}}$.
{  In addition to the spatial distribution, we present the prompt and ICS emission spectra in the right panel of Fig. \ref{fig:spectrum}. The ICS radiation component is stronger than the prompt emission component over a wide energy range.}
As the spherical harmonics analysis can reveal detailed information about the physical quantity distributed on the surface of a sphere, we store the count map in HEALPIX \footnote{\url{http://healpix.sourceforge.net/}} form (with resolution parameter NSIDE=512) as well in Sec. \ref{sec:IIIB}, in order to carry out the analysis.

GC hosts a lot of sources with energetic activities that could be accelerators of cosmic rays.
Therefore GC is not a perfect region to analyze the DM origin signal from an observational aspect. For this reason, we also consider the case of masking the regions of $|b|<2^\circ$  in our analysis.

\begin{figure}
    \begin{center}
        \includegraphics[width=0.7\linewidth]{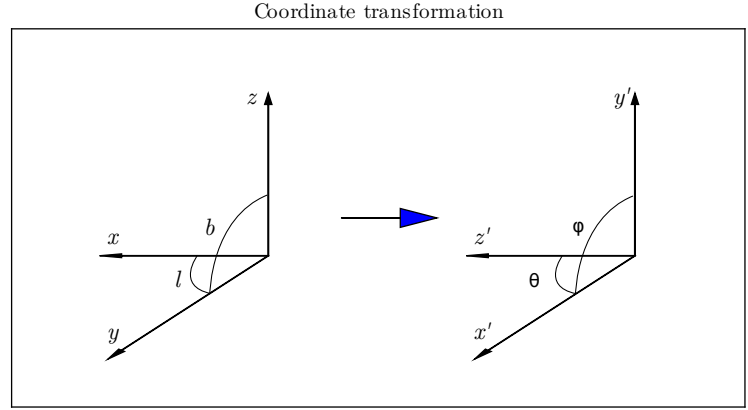}
    \end{center}
    \caption{The coordinate transformation. In the left panel, the $z$-axis is perpendicular to the Galactic plane in the Galactic coordinate system. In order to better reveal the rotational asymmetry we choose a new coordinate system (i.e., the right panel) in which the new $z$-axis is the original $x$-axis pointing from sun to the GC.}
    \label{fig:coor_trans}
\end{figure}

It is well known that the spherically symmetric emission about the GC will be rotationally symmetric for observers on the Earth.
In other words, the rotational symmetry along our line of sight to the GC reflects the spherical symmetry with respect to GC.
In order to analyze the rotational asymmetry, we examine the variance of the flux in a given viewing angle region (i.e., $\theta_1<\theta<\theta_2$, where $\cos\theta=\cos l \cos b$) but at different $\varphi$.
A coordinate transformation (see Fig. \ref{fig:coor_trans}) is therefore needed to manifest the asymmetry. The equations of coordinate transformation are
\begin{equation}
\begin{aligned}
  \sin  b =\sin \theta  \sin \varphi,~~~\tan l = \tan \theta \cos \varphi.
\end{aligned}
\end{equation}
\begin{figure}
    \begin{center}
        \includegraphics[width=0.45\linewidth]{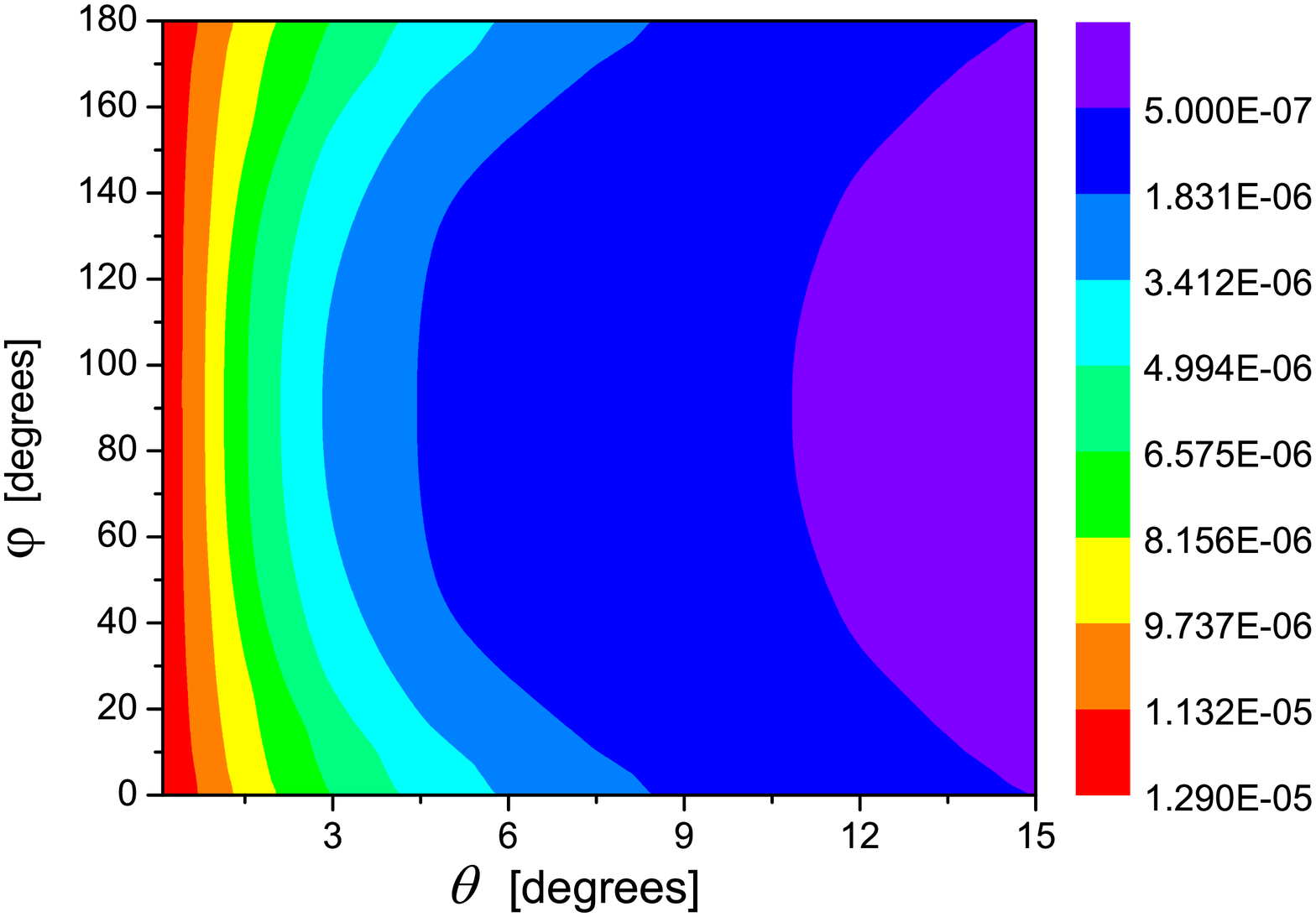}
       \includegraphics[width=0.45\linewidth]{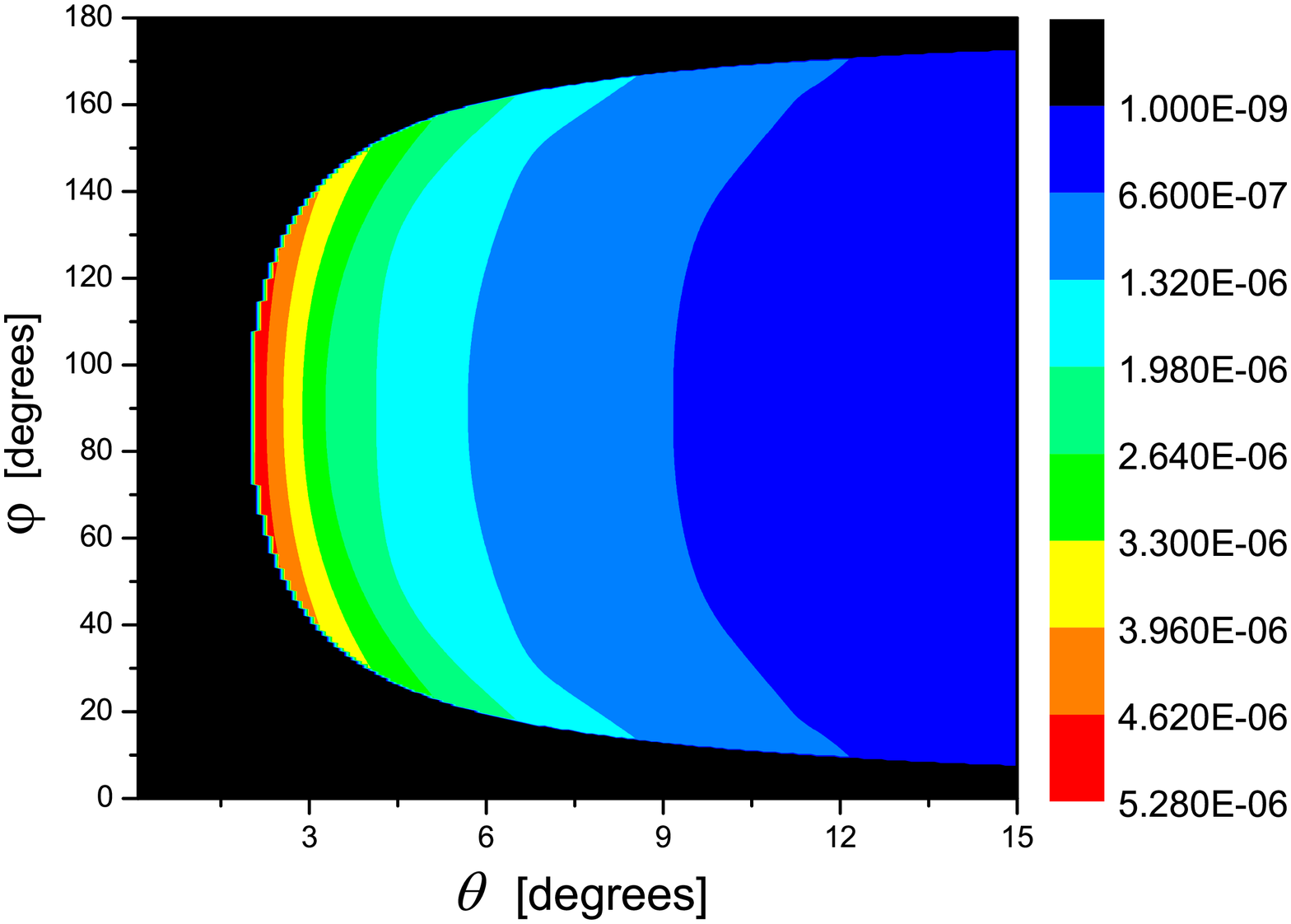}
    \end{center}
    \caption{The spatial distribution of the ICS emission viewed in the coordinate system of $(\theta,~\varphi)$. The figures above are offered as an illustration of rotational asymmetry of two count maps in Fig. \ref{fig:countmap}.}
    \label{fig:show}
\end{figure}
In the new coordinate system, spherical coordinate $\theta$ is the angle observed away from the GC. At a given $\theta$, the flux of gamma-ray emission varying with spherical coordinate $\varphi$ reflects the asymmetry.
In Fig. \ref{fig:show}, we show part of the results of coordinate transformation of two count maps in Fig. \ref{fig:countmap}, where $\theta$ and $\varphi$ are set as the horizontal and longitudinal coordinates  respectively. The change of the flux along the longitudinal coordinate manifests the rotational asymmetry. As shown in both the left and the right panels of Fig. \ref{fig:show}, non-negligible rotational asymmetry displays in the data, which is independent of whether or not we mask the regions of $|b|<2^{\circ}$.

\begin{figure}
    \begin{center}
        \includegraphics[width=0.38\linewidth]{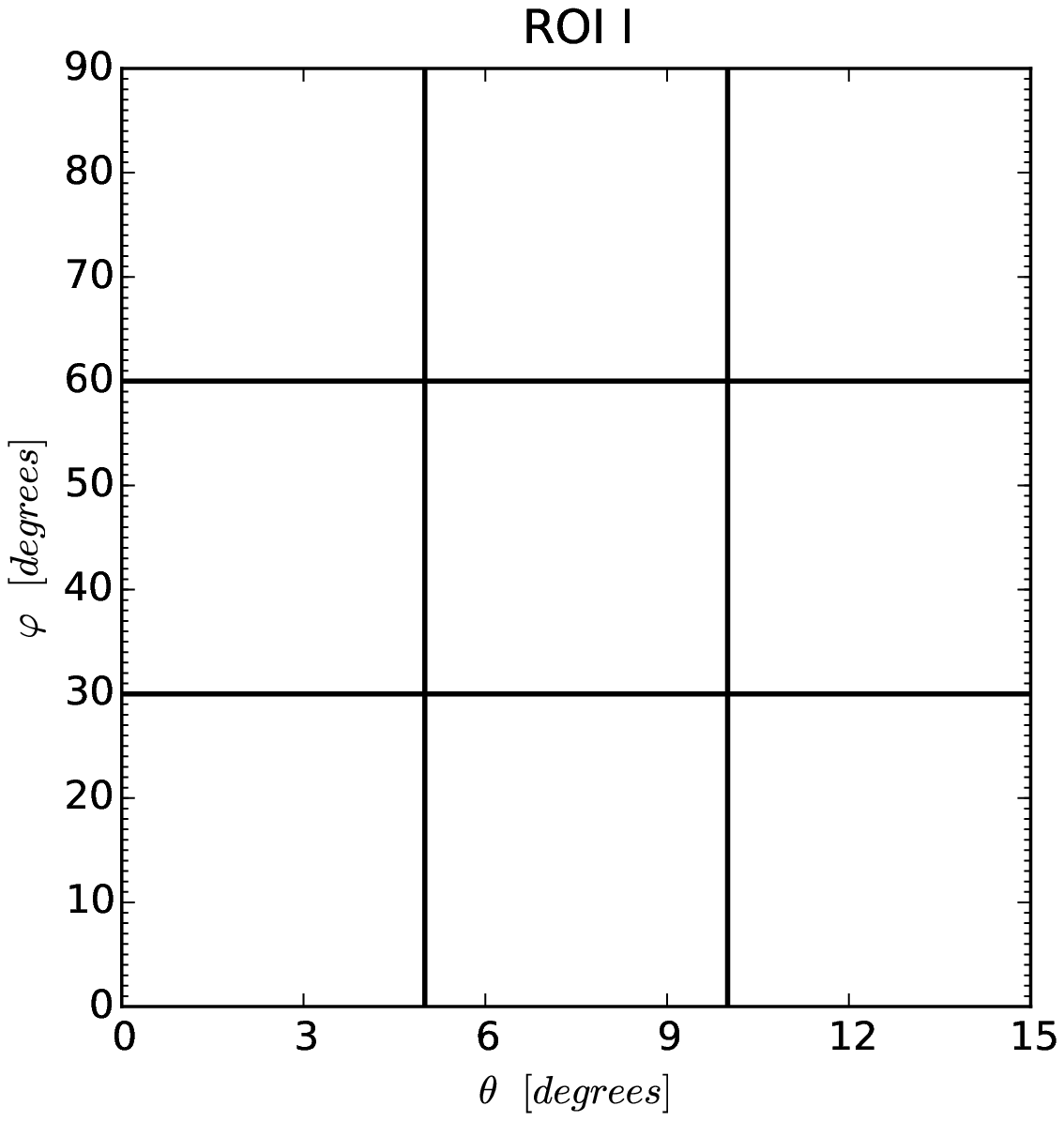}
        \includegraphics[width=0.38\linewidth]{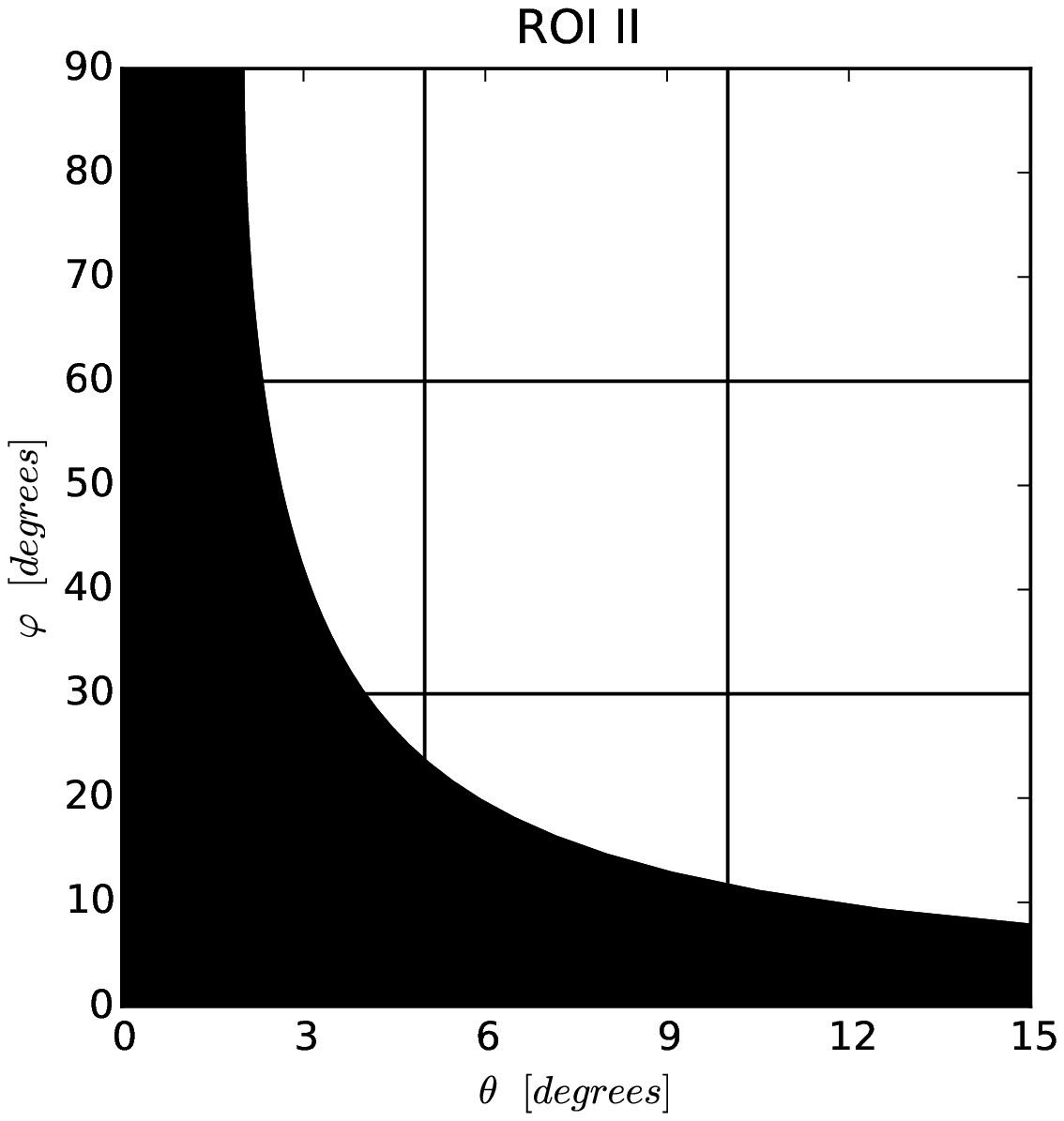}
    \end{center}
    \caption{The regions of interest adopted in estimating the DOA of the spatial distribution of diffuse $\gamma$-ray emission. The left panel is for ROI I (i.e., $b>0^\circ$, $l>0^\circ$, and $0^\circ<\varphi<90^{\circ}$). The right panel is for ROI II, which is different from ROI I by masking  $b<2^\circ$.}
    \label{fig:skysegment}
\end{figure}

Now we optimize our ROIs to reduce the calculation time as well as the uncertainties for possible observational test in the future.
The field of starlight is not only approximately axial symmetric about the GC but also mirror symmetric about the Galactic plane. So 1/4 of the count map in Fig. \ref{fig:countmap} is enough if one is only interested in the ICS emission. Therefore we select a ROI in Sec. \ref{sec:IIIA} as $0^\circ<\theta<15^\circ$ and $0^\circ<\varphi<90^\circ$ (i.e., ROI I),
as shown in the left panel of Fig. \ref{fig:skysegment}.
Considering the complication about GC, we mask the region $b<2^\circ$ of ROI I and denote it as ROI II (i.e. right panel of Fig. \ref{fig:skysegment}) for further study.
In Sec. \ref{sec:IIIB}, however, we analyze the whole celestial sphere in order to conduct the spherical harmonics analysis. The regions of the whole sky with and without ($|b|<2^\circ$) are defined as ROI III (see the left of top and middle panel of Fig. \ref{fig:aps}) and  ROI IV  ( see left of bottom panel of Fig. \ref{fig:aps}), respectively. While in Sec. \ref{sec:IIIC}, the ROI V (i.e. the regions of $b>2^\circ$, $0^\circ<\theta<15^\circ$ and $0^\circ<\varphi<360^\circ$) is slightly different from ROI II that we don't show specifically. The reason is that the distribution of gas in the Milky Way doesn't share the approximate symmetry with starlight.

\section{Studies of the morphology of the $\gamma-$ray emission}\label{Sec:analysis}

\subsection{Degree of asymmetry of $\gamma$-ray emission}\label{sec:IIIA}
In this section, we analyze the asymmetry in the new coordinate system (see Fig. \ref{fig:coor_trans}). The ICS emission flux varying with $\varphi$ directly reflects the asymmetry at a given spherical coordinate $\theta$. We prefer the flux data on average in several sub-regions instead of analyzing value of special points in case of statistical errors for the observational test.
Hus, we divide the ROI I into three sub-regions (i.e., $0^\circ<\theta<5^\circ$, $5^\circ<\theta<10^\circ$ and $10^\circ<\theta<15^\circ$) and split each sub-region equally into three segments. The ROI II is divided in the same way as for ROI I except that the three segments are no longer equal since the regions of $|b|<2^\circ$ have been masked. Additionally, one segment ($0^\circ<\theta<5^\circ$, $b>2^\circ$, $0^\circ<\varphi<30^\circ$) is so small that it can be abandoned. Comparing the variance of average flux of these segments in a sub-region, we could get the maximum, the minimum and the mean value of these segments (i.e., $F_{\rm max},~F_{\rm min},~F_{\rm mean}$). Let us define the degree of asymmetry (DOA) in order to give a quantitative description of the spatial distribution, i.e.,
\begin{equation}
\begin{aligned}
{\rm DOA}\equiv (F_{\rm max}-F_{\rm min})/F_{\rm mean}.
\end{aligned}
\end{equation}

Table \ref{ICSresult} presents our DOAs in the sub-regions of ROI I and ROI II, respectively. Significant asymmetries (i.e., DOA $>10\%$) are found in most sub-regions. The values of DOA in ROI I  are found to be larger than those in ROI II. The reason is that the flux in the Galactic plane is larger than in other places.
 If the $b<2^\circ$ regions are masked, the average flux of the segments that contain the masked area decreases. As a result, the change of average flux is ``suppressed".
In another word, the mask of some sky along the Galactic plane would reduce the DOA.
\begin{table*}
\begin{center}
\caption{The DOA of ICS emission in ROI I and ROI II}
\begin{tabular}{cccccc}
\hline\hline
 sub-regions of ROI I & DOA  &   &   & Sub-regions of ROI II & DOA  \\
\hline
$b>0^\circ $,  $0^\circ<\theta<5^{\circ}$     &   $0.48$    &     &    &$b>2^\circ $,  $0^\circ<\theta<5^{\circ}$     &   $0.06$\\
\hline
 $b>0^\circ $,  $5^\circ<\theta<10^{\circ}$   &    $0.66$  &    &    & $b>2^\circ $,  $5^\circ<\theta<10^{\circ}$   &    $0.36$ \\
\hline
 $b>0^\circ $,  $10^\circ<\theta<15^{\circ}$ &    $0.46$  &    &    &  $b>2^\circ $,  $10^\circ<\theta<15^{\circ}$ &    $0.34$\\
\hline
\end{tabular}
\label{ICSresult}
\end{center}
\end{table*}


\subsection{Spherical harmonics expansion}\label{sec:IIIB}
As for cosmic microwave background (CMB), we carry out a spherical harmonics expansion to spatial distribution of $\gamma$-ray induced by DM annihilations despite having limited data \cite{spherana}.
Since a set of fundamental cosmological parameters have been inferred from angular power spectrum (APS) of the CMB \cite{Ade:2013aa}, the analysis of the APS of spatial distribution of the possible DM-induced $\gamma$-ray emission may yield more information on its nature.

\begin{figure}
    \begin{center}
        \includegraphics[width=0.42\linewidth]{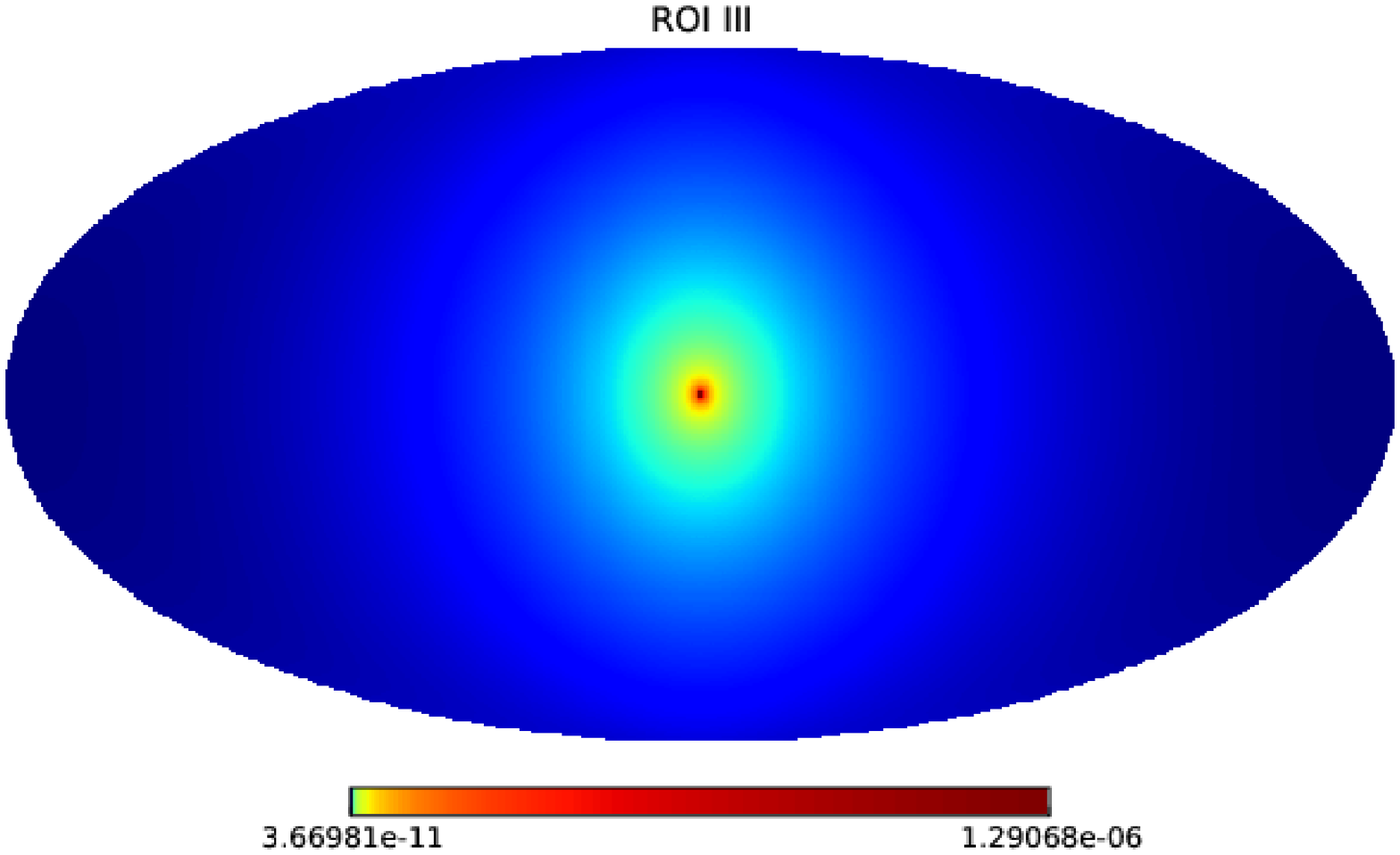}
        \includegraphics[width=0.42\linewidth]{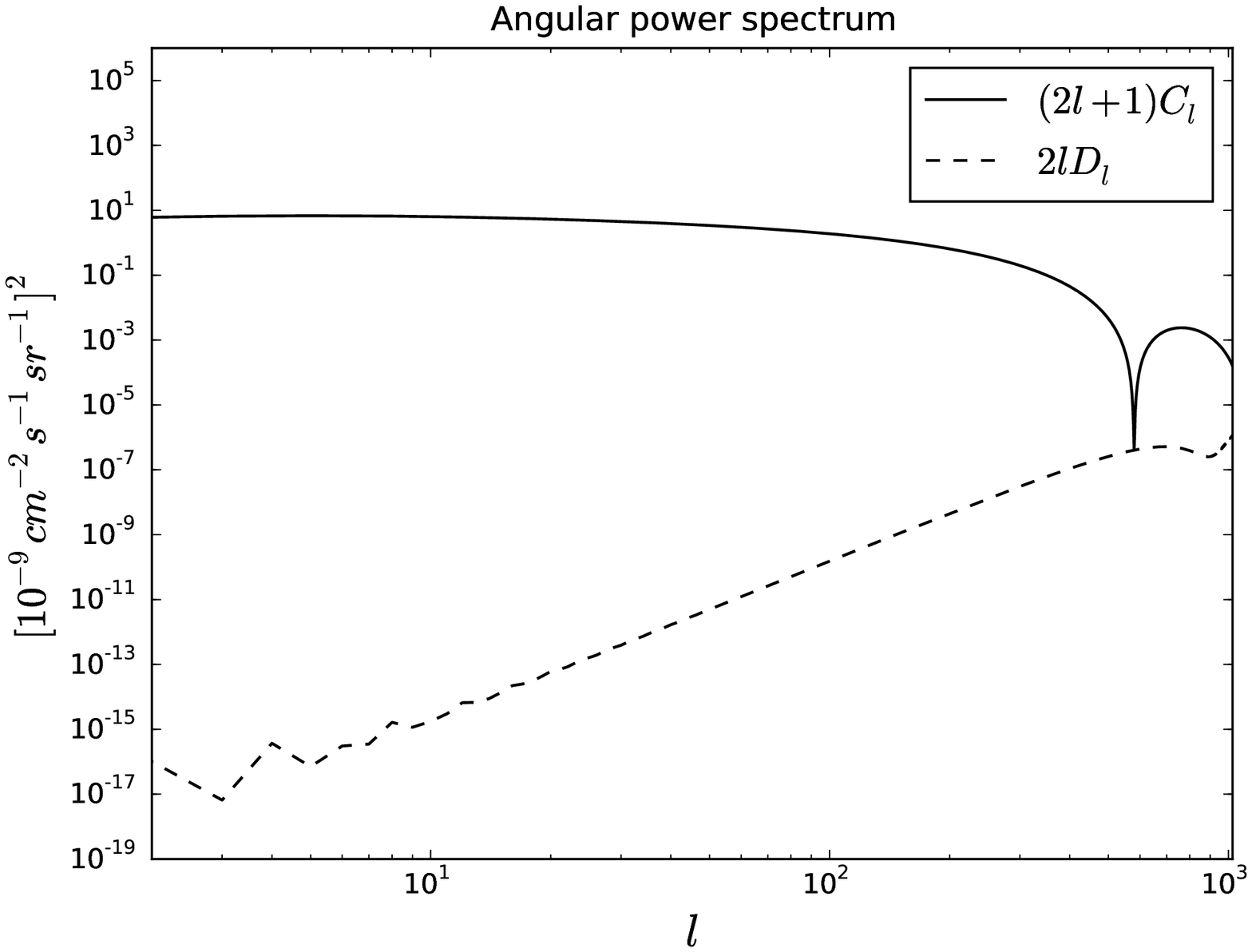}   \\
        \includegraphics[width=0.42\linewidth]{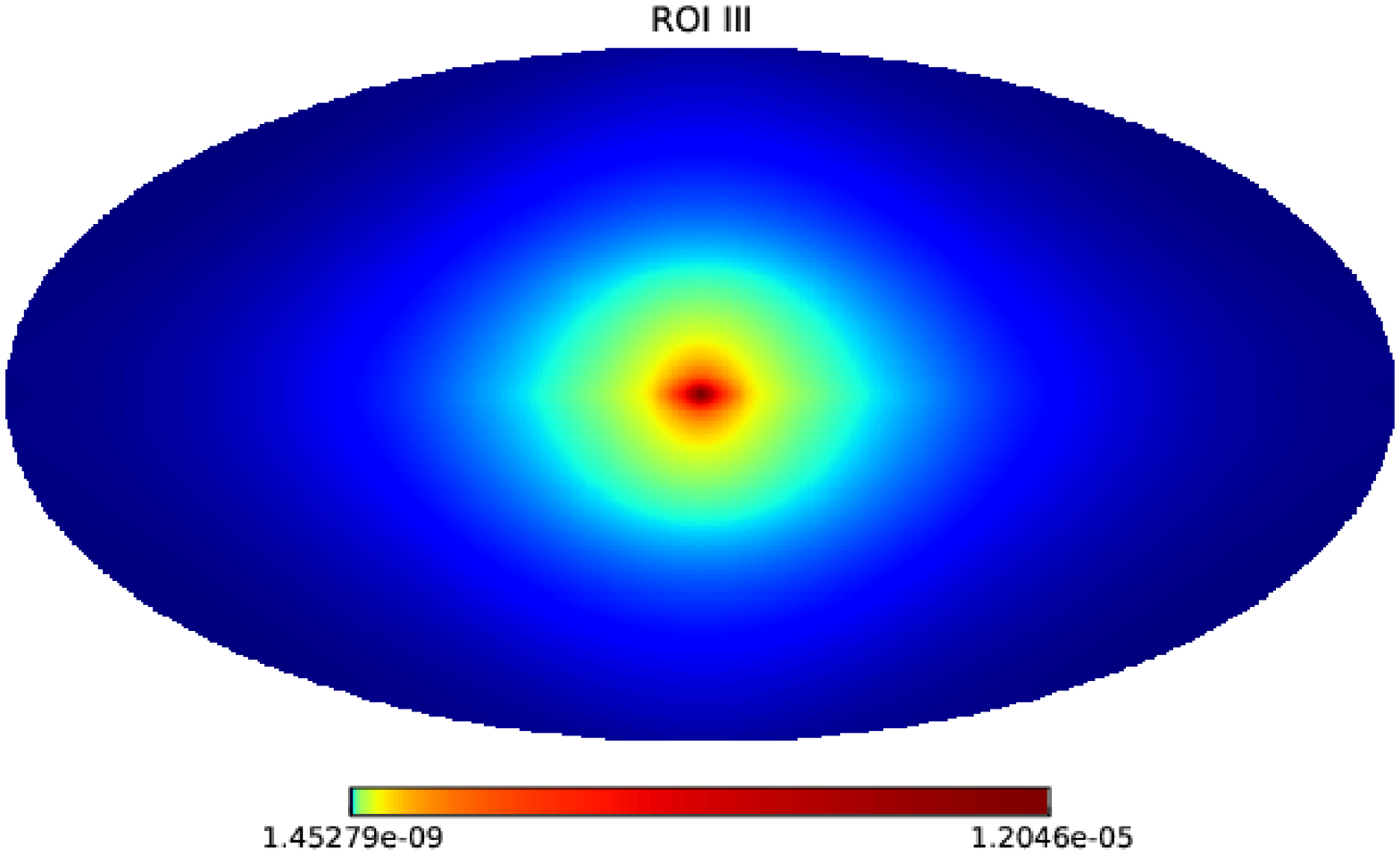}
       \includegraphics[width=0.42\linewidth]{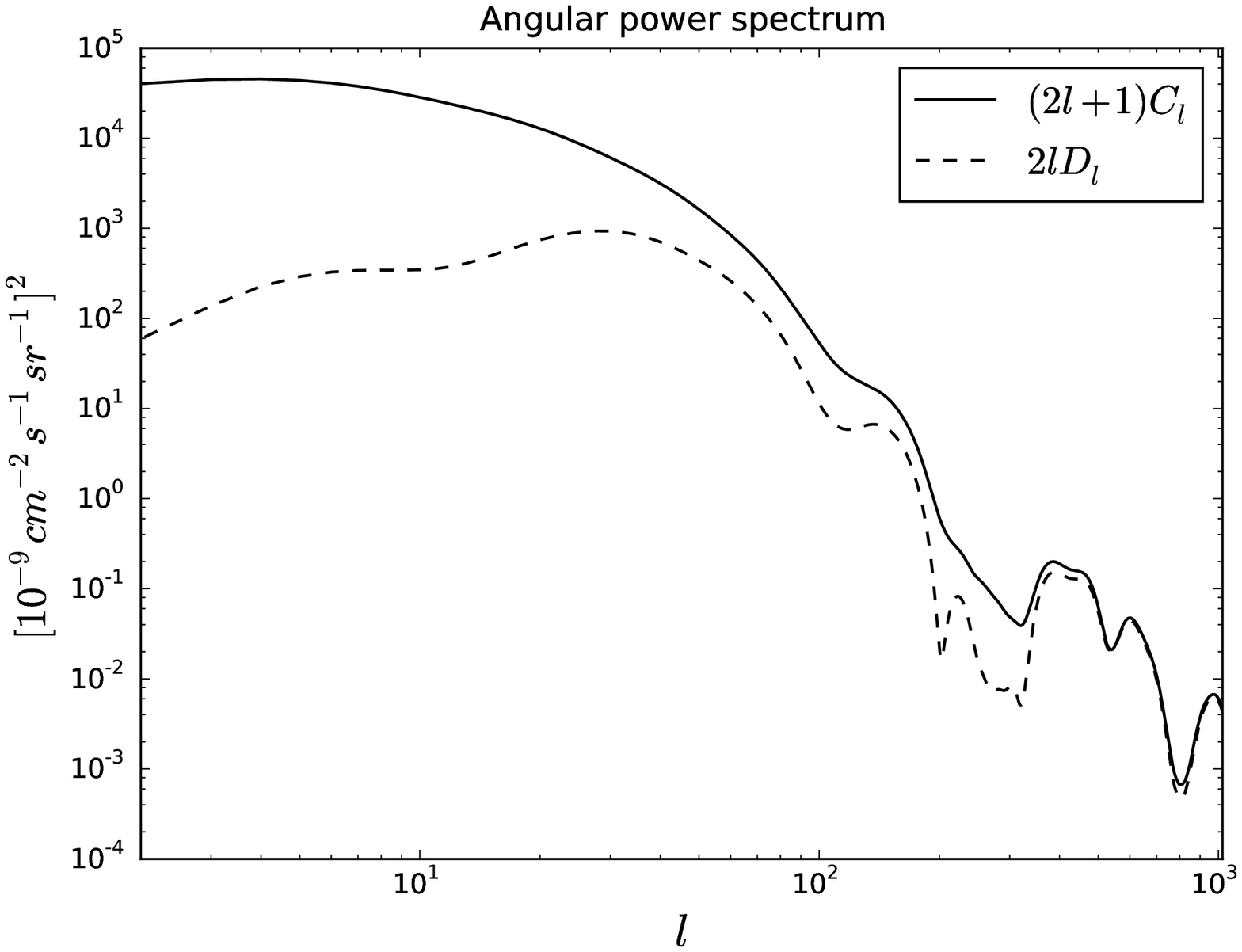}   \\
       \includegraphics[width=0.42\linewidth]{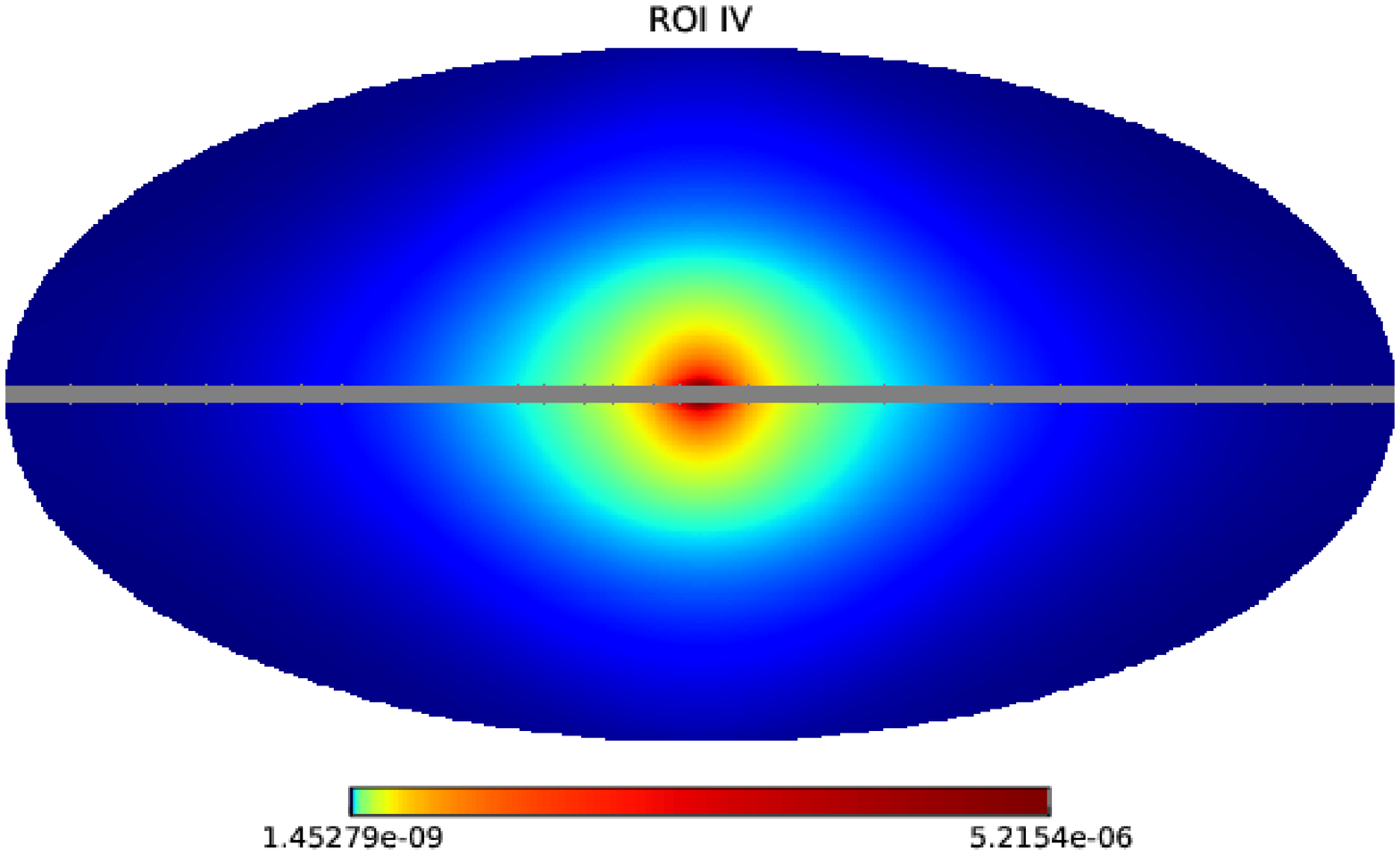}
        \includegraphics[width=0.42\linewidth]{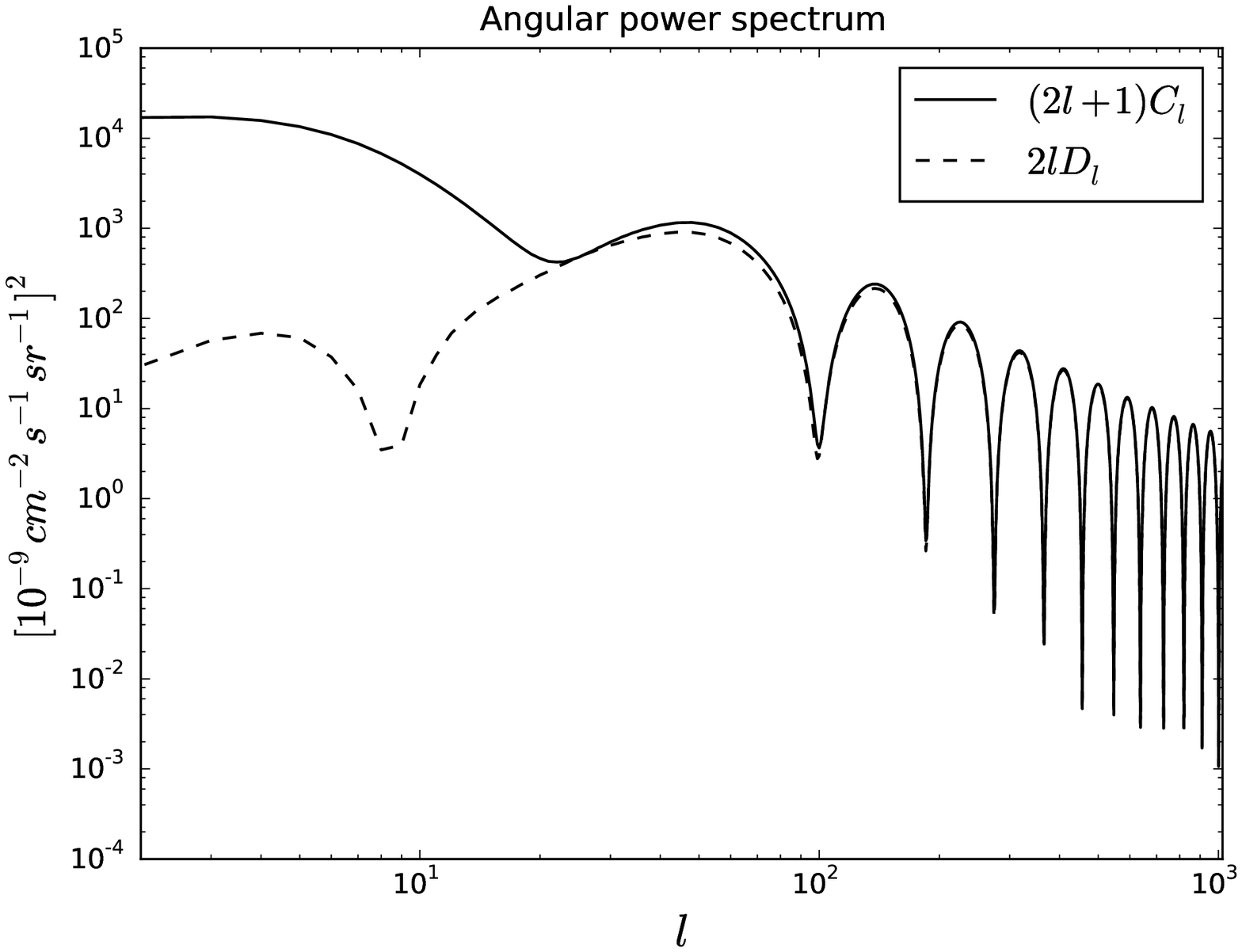}
    \end{center}
    \caption{
The left column presents the count maps of $\gamma$ rays in logarithmic scale in Galactic coordinate system. And the two kinds of APS illustrated in the right column are in the new coordinate system: the solid line represents $C_{l}$ while the dotted line stands for $D_{l}$. We make the maps of count to HEALPIX grids with NSIDE=512 and  expand it to $l_{max}$=1024.
Top panel shows results in ROI III for prompt emission of SIDM annihilation while middle panel is for ICS. The bottom panel is same as middle panel except in ROI IV.  }
    \label{fig:aps}
\end{figure}

In Sec. \ref{Sec:calculation} we have mentioned that spherical harmonics analysis would be done in the new coordinate system (see Fig. \ref{fig:coor_trans}).
Using HEALPIX, we analyze the count map of $\gamma$ rays in the energy range of $1~{\rm GeV}-3.16~{\rm GeV}$ via spherical harmonics expansion. If some coefficients of $Y_{lm}$ terms are nonzero for ${\rm m} \neq 0$, there should be departure from the rotational symmetry. The APS $C_{l}$ is defined as  $C_{l}$ = $\sum |a_{lm}|^{2}/(2l+1) $, where  $|{\rm m}|\leq l$. In order to better show the asymmetries, we take $D_{l}$ = $\sum |a_{lm}|^{2}/2l$, where ${\rm m} \neq 0$.
In order to check the approach, we analyze the count map of prompt $\gamma$-ray emission of channel: $\bar{\chi}\chi\rightarrow \phi\phi\rightarrow e^{+}e^{-}e^{+}e^{-}$, which is rotationally-symmetric. The results are shown in top panel of Fig. \ref{fig:aps}, where count map in logarithmic scale is on the left  and two types of APS are on the right. It's clear that the coefficients of $Y_{lm}$ terms for ${\rm m} \neq 0$ are relatively  negligible for most of multiple $l$. However, the method is not valid when $l$ is approaching 1024, due to numerical errors.

Two kinds of power spectra are shown in the right of Fig. \ref{fig:aps} while the figures on the left are count maps in logarithmic scale. The APS without masking the regions of $|b|<2^{\circ}$ (see  the right of middle panel of Fig. \ref{fig:aps}) is smooth and easily analyzed. For $l<30$, the asymmetry is negligible as the value of $D_{l}$ is much smaller than that of $C_{l}$.
When $l$ reaches ~ 50, the asymmetry becomes important.
When masking the regions of $|b|<2^{\circ}$, the APS oscillates quickly (see the right of bottom panel of Fig. \ref{fig:aps}).
{  The oscillation of APS is due to masking of such regions and some DOA information is lost. 
As a result, it's not proper to analyze qualitatively  when the APS oscillates (e.g. $l>30$).}

\subsection{The robustness of the DOA}\label{sec:IIIC}
In the previous studies of DOA, we don't take the bremsstrahlung and prompt emission components into account since they are relatively weak compared to the ICS for SIDM annihilation channel \cite{PPPC}. The bremsstrahlung diffuse emission flux in the Galactic plane would be much larger than in other places since the Galactic plane contains dense gas. Hence the rotational asymmetry of the bremsstrahlung component is expected to be larger than that of the ICS component. The DOA increases if both the bremsstrahlung and ICS components are considered together. As for the prompt emission component, we utilize the PPPC4DMID \cite{PPPC} to calculate the spectra of the DM annihilation. We also adopt the same generalized NFW profile, as
was used in Sec. \ref{Sec:calculation}. Contrary to bremsstrahlung emission, the prompt emission of DM annihilation will not give rise to a DOA.
If these three components are summed, the net DOA would be determined by the relative amounts of these constituents. Since bremsstrahlung emission doesn't possess the symmetry of ICS, we compute the DOA in ROI V instead of ROI II to obtain a more precise result.
The results of DOAs are summarized in Table \ref{sum_mask}.
Comparing with the DOA of ICS presented in the second column of Table \ref{ICSresult}, we find that at least for our fiducial parameters the bremsstrahlung and prompt emission components do not play an important role in modifying the DOA.

\begin{table*}
\begin{center}
\caption{DOA of the total emission of ICS, bremsstrahlung and prompt in ROI V}
\begin{tabular}{cc}
\hline\hline
Sub-regions of ROI V & DOA  \\
\hline
$b>2^\circ $,  $0^\circ<\theta<5^{\circ}$     &   $0.06$  \\
\hline
 $b>2^\circ $,  $5^\circ<\theta<10^{\circ}$   &    $0.37$  \\
\hline
 $b>2^\circ $,  $10^\circ<\theta<15^{\circ}$ &    $0.35$  \\
\hline
\end{tabular}
\label{sum_mask}
\end{center}
\end{table*}
Let us also check whether or not the high anisotropy of the diffuse emission originating from SIDM might be a consequence of utilizing a special group of cosmic ray propagation parameters. Such a possibility can be tested by varying the propagation parameters. We re-calculate the DOA by adding ICS, Bremsstrahlung and Prompt emission components in ROI V using two groups of parameters listed in Table \ref{pram} \cite{FermiLAT:2012aa}. Such a range including the parameters used in previous studies will largely cover the uncertainties of the propagation parameters \cite{Trotta_CR,Jin_CR}. We would refer to the two rows of Table \ref{pram} as parameter set 1, parameter set 2 respectively. The results are depicted in Table \ref{sum_param}. Despite suffering from some variations compared with the DOA listed in Table \ref{sum_mask}, the DOAs are still significant, suggesting that our previous conclusions are robust.\\

\begin{table*}
\begin{center}
\caption{The adopted two sets of cosmic ray propagation parameters.}
\begin{tabular}{ccccccccc}
\hline\hline
  &  &  $D_{0}$       &                               &   $z_{h}$    &    & $v_{A}$   &    & $\delta$  \\
  &  &  ($10^{28}~\rm cm^{2}~s^{-1}$) &   &   (kpc)         &    & (${\rm km~s^{-1}}$)&  & \\
\hline
     Set 1 &   &  2.7 &    &  2   &   & 35.0  &   &  0.33   \\
\hline
    Set 2  &   &  9.4  &    & 10  &   & 28.6  &   &   0.33   \\
\hline
\end{tabular}
\label{pram}
\end{center}
\end{table*}

\begin{table*}
\begin{center}
\caption{The DOAs (in ROI V) obtained for different sets of cosmic ray propagation parameters.}
\begin{tabular}{ccccc}
\hline\hline
 Sub-regions of ROI V  &                                                   & Set 1 &   &   Set 2 \\
$b>2^\circ $,  $0^\circ<\theta<5^{\circ}$  &       &   0.03 &   & 0.07  \\
\hline
 $b>2^\circ $,  $5^\circ<\theta<10^{\circ}$ &    &   0.39 &   & 0.39  \\
\hline
 $b>2^\circ $,  $10^\circ<\theta<15^{\circ}$ &     &  0.86 &   & 0.37  \\
\hline
\end{tabular}
\label{sum_param}
\end{center}
\end{table*}

However, there could be an exception if the dark matter particles have a $m_\chi\sim 1$ TeV. Though in this work we focus on $m_\chi \sim $tens GeV that might be favored by the GCE emission, for completeness here we also briefly examine the case of $m_\chi\sim 1$ TeV (see \cite{zhangjuan} for the case of $\chi\chi\rightarrow e^{+}e^{-}$), for which the ICS emission component may have a much smaller DOA. The reason is that for $m_\chi \sim $tens GeV, the resulting electron/positron pairs have $\gamma_{\rm e^{\pm}}\sim 10^{4}-10^{5}$ and they mainly scatter with the starlight. While for $m_\chi \sim 1$TeV, the formed electron/positron pairs have $\gamma_{\rm e^{\pm}}\sim 10^{6}$. In the rest frame of such energetic particles, the starlight has an energy $\approx 10^{6}~{\rm eV}(\gamma_{\rm e^{\pm}}/10^{6})(\epsilon_{\rm starlight}/1~{\rm eV})$, exceeding the rest mass of the electrons/positrons. As a result, the inverse Compton scattering is effectively suppressed.
 Instead, the TeV electrons mainly scatter the CMB and infrared photons and can boost some of them to $\sim 1~{\rm GeV}~(\gamma_{\rm e^{\pm}}/10^{6})^{2}$, again well within the energy range of the GCE emission.
We calculate the DOA in ROI V.
There is no sizeable DOA in all sub-regions (less than 10$\%$).

\section{Discussion}\label{Sec:discussion}
With the Fermi-LAT data quite a few research groups have reported a very-significant spatially extended GeV $\gamma$-ray excess surrounding the Galactic Center. The physical origin of such a GeV excess is highly debatable and an interesting possibility is the ICS of the electrons/positrons from annihilation of self-interacting dark matter particles with the interstellar optical photons. In such a scenario, the constraints set by the non-detection of a clear signal in the dwarf spherical galaxies is likely non-applicable since the $e^\pm$ pairs produced via dark matter annihilation will not produce plentiful $\gamma$ rays due to the low starlight/gas-densities. Motivated by such facts, in this work we have investigated morphology of the GeV $\gamma$-ray emission resulting in the ICS process.
The regions of $|b|<15^\circ$ have been explored and the DM density distribution has been taken as the generalized NFW profile with $r_{\rm s}=20$ kpc and the slope index $\alpha=1.2$. The annihilation channel of $\bar{\chi}\chi\rightarrow \phi\phi\rightarrow e^{+}e^{-}e^{+}e^{-}$ has been investigated. The general conclusion is that for $\theta>5^{\circ}$, the degree of rotational asymmetry reaches $30\%$ or even larger for $m_\chi\sim$ tens GeV, independent of the propagation parameters. The physical reason is that for the tens of GeV electrons/positrons, the cooling is not quick enough to lose significant portions of their kinetic energy locally. As a result, the ICS emission traces the distribution of the starlight, which is expected to be most dense along the Galactic plane since most of the stars concentrate in such a region.
For the same reason, though in this work we have only discussed the ICS process of the tens GeV electrons/positrons from dark matter annihilation, a significant DOA is expected for the ICS GeV emission of the tens GeV electrons/positrons originating from other astrophysical processes (for example, the millisecond pulsars are also believed to be high energy electron/positron sources. We note that such a scenario has been investigated recently by \cite{Lacroix:2015}). One caution is that, if instead the electron/positron pairs were from the annihilation of dark matter particles with a $m_\chi \gtrsim 1$ TeV, the DOA seems negligible. 

The sizeable DOA found in the regions of $\theta>5^\circ$ is helpful in testing the SIDM ICS interpretation of the GeV excess.
As found in the latest analysis by the Fermi collaboration \cite{FermiLAT:2015}, if only interstellar
emission and point sources are fit to the data the residual GeV emission
is weakly asymmetric about the GC, but the statistical noise is large. This may be suggestive of an excess in the data
that is not symmetric with respect to the GC. {  However, the current astrophysical background gamma-ray emission in particular in the direction of the Galactic center is still to be better constrained. Different Galactic diffuse emission models yield different GCE spectra and the difference can be up to $\sim30\%$ or even larger (e.g., \cite{Calore:2014xka, Zhou:2014lva}). Fortunately, the Fermi-LAT team is developing a new model of Galactic diffuse emission, with which the uncertainties of astrophysical background can be significantly reduced \cite{Johannesson:2015}.} With the improved diffuse background model a reliable asymmetry of the GCE emission signal is expected, which can then be used to reliably test the ICS interpretation of GCE within the scenario of SIDM annihilation.

\acknowledgments We thank Q. Yuan, L. Feng, X. Li, P.-F. Zhang and Z.-Q. Shen for helpful discussions. This work is supported in part by the National Natural Science Foundation of China (under Grants No. 11275097, No. 11475085, and No. 11535005).

\bibliographystyle{apsrev}
\bibliography{refs}

\end{document}